\documentclass [11pt, letterpaper]{article}
\pdfoutput=1 
\usepackage{fullpage}
\usepackage{amsmath}
\usepackage{amssymb}
\usepackage{graphicx}
\usepackage{mathrsfs}
\usepackage{wrapfig}
\usepackage{boxedminipage}
\usepackage{epsfig}
\usepackage{setspace}
\usepackage{subfigure}
\usepackage{float}
\usepackage{hyperref}
\usepackage{enumerate}
\usepackage{cite}
\usepackage[font=footnotesize,labelfont=rm]{caption}

\begin{document}

\begin{flushright}
{\tt arXiv:1705.04855}
\end{flushright}

{\flushleft\vskip-1.35cm\vbox{\includegraphics[width=1.25in]{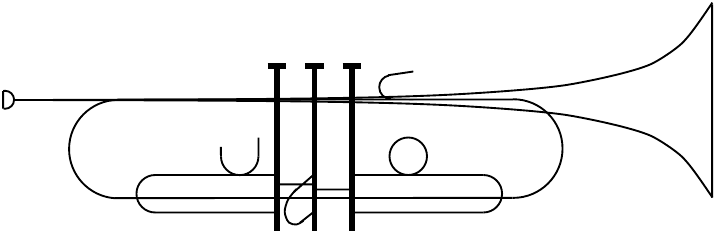}}}

\bigskip
\bigskip
\bigskip
\bigskip

\bigskip
\bigskip
\bigskip
\bigskip

\begin{center} 

{\Large\bf  Taub--Bolt  Heat Engines}

\end{center}

\bigskip \bigskip \bigskip \bigskip

\centerline{\bf Clifford V. Johnson}

\bigskip
\bigskip
\bigskip

  \centerline{\it Department of Physics and Astronomy }
\centerline{\it University of
Southern California}
\centerline{\it Los Angeles, CA 90089-0484, U.S.A.}

\bigskip

\centerline{\small \tt johnson1,  [at] usc.edu}

\bigskip
\bigskip

\begin{abstract}
It is shown that aspects of the  extended thermodynamic properties of the Taub--Bolt--AdS spacetime in four dimensions are  similar to those of the Schwarzschild--AdS black hole.  In a high temperature expansion, the equations of state begin to deviate only at next--to--subleading orders. By analogy with what has been done for black holes, Taub--Bolt's thermodynamic equations are  used to define  holographic heat engines,  the first examples of gravitational heat engines defined using a spacetime that is not a black hole. As a further comparison, the Taub--Bolt engine efficiency  is computed for two special kinds of engine cycle and compared to the results for  analogous Schwarzschild black hole engine cycles.
\end{abstract}
\newpage \baselineskip=18pt \setcounter{footnote}{0}


\section{Introduction}
\label{sec:one}
The extended  thermodynamics arising from the semi--classical quantum treatment of gravitational systems~\cite{Bekenstein:1973ur,Bekenstein:1974ax,Hawking:1974sw,Hawking:1976de,Gibbons:1976ue} (in which there is additionally\footnote{See {\it e.g.,} refs.~\cite{Wang:2006eb,Sekiwa:2006qj,LarranagaRubio:2007ut,Kastor:2009wy},  the early work in refs.~\cite{Henneaux:1984ji,Teitelboim:1985dp,Henneaux:1989zc}} a pressure $p$ and volume $V$) can produce interesting equations of state. These  can be used as  valuable thermodynamic models beyond the gravitational context in which they might first arise. Part of the attraction  here  is that explicit knowledge of   the underlying spacetime metric  can produce closed form expressions for several thermodynamic quantities even in regimes where standard computational tools from statistical mechanics and thermodynamics might be difficult to apply. Although it is still too early to tell, this could be a toolbox with  wide applications\footnote{See for example ref.~\cite{Johnson:2017hxu}.}.

While most of the focus of studies of extended thermodynamics has been on black hole spacetimes and the rich family of phenomena that can be explored using them\footnote{See the reviews in {\it e.g.,} refs. \cite{Dolan:2012jh,Kubiznak:2016qmn} for a review of some of the  results obtained  in this context by using the extended thermodynamics framework.}, it was pointed out in ref.~\cite{Johnson:2014xza}  that the core extended thermodynamics quantities ({\it e.g.} the enthalpy $H$ and thermodynamic volume~$V$) can be computed and given meaning for spacetimes that are not black holes. The examples used in that work were the Taub--NUT and Taub--Bolt spacetimes, (both for negative cosmological constant~$\Lambda$, and for $\Lambda=0$ as  a smooth limit). A special case of the Taub--NUT example also shows that it works for a special time--slicing of pure AdS$_4$ \cite{Johnson:2014xza}, and a recent paper \cite{Johnson:2017asf}  defined the extended thermodynamic quantities for flat space Rindler in $D$ dimensions by studying a special double limit (large--charge and near--horizon) of a family of critical charged black holes.

Putting aside whether the various Euclideanized spacetimes that might be treated actually have a Lorentzian interpretation or not, the thermodynamic quantities and equations of state derived from them can be taken at face value as interesting model systems to study. They may one day perhaps have physical applications, although such applications are currently unknown.

This paper begins by demonstrating that the extended thermodynamic quantities derived in ref.~\cite{Johnson:2014xza} for Taub--Bolt--AdS$_4$ (building on the traditional thermodynamic work of ref.~\cite{Hunter:1998qe,Chamblin:1998pz,Hawking:1998ct}) together define a system that, looked at the right way, is (locally) very similar to the Schwarzschild--AdS$_4$ system of thermodynamics \cite{Hawking:1982dh,Kastor:2009wy,Dolan:2010ha}. Specifically, they have the same behaviour at leading orders in a high temperature expansion, but begin to deviate from each other at subleading orders.  

The two spacetimes can be readily seen to differ globally, of course since the ``NUT--charge" parameter (see below)  comes with a topological twist, making the Euclidean boundary at large radius not $S^1\times S^2$, but $S^3$. However,  in the spirit of the opening paragraphs, at face value the thermodynamic  systems are {\it locally} extremely similar, as will be explored below.

One way to compare two thermodynamic systems is to put them to work: Black holes can be used to define ``holographic heat engines" \cite{Johnson:2014yja}---cyclic systems that, with a net heat input $Q_H$ and output $Q_C$ per cycle, do mechanical work {\it via} the $pdV$ terms in the First Law of thermodynamics:
\begin{equation}
\label{eq:FirstLawA}
dU=TdS-pdV\ .
\end{equation}
The other main point of this paper is to observe that the thermodynamic equations of Taub--Bolt can also be used to define  heat engines. This is likely to be the case for other non--black hole spacetimes.  We compute the efficiency $\eta=W/Q_H=1-Q_C/Q_H$ for Taub--Bolt and Schwarzschild engines for two different kinds of engine cycle in order to compare the two classes of engine.  The comparison has an extra wrinkle of interest to it since the expressions for the thermodynamic volume and entropy of the Taub--Bolt spacetime are non--trivial (as compared to Schwarzschild), and the specific heat at constant volume ($C_V$) is non--vanishing, in contrast to most examples studied in the literature so far\footnote{For the only other $C_V\neq0$ examples of holographic heat engines so far, see the recent work in refs.~\cite{Hennigar:2017apu,Chakraborty:2017weq}.}.

\section{Comparison of Thermodynamic Properties}
\label{sec:two}
Let us  work in  $D=4$, with action\footnote{Here, the conventions of refs.~\cite{Chamblin:1999tk,Chamblin:1999hg} will be used.}:
\begin{equation}
I=-\frac{1}{16\pi }\int \! d^4x \sqrt{-g} \left(R-2\Lambda -F^2\right)\ .
\label{eq:action}
\end{equation}
  The cosmological constant $\Lambda=-3/l^2$  sets a length scale~$l$, and we have set Newton's constant $G$ and the speed of light $c$ to unity (as we will later for~$\hbar$ and~$k_B$). The spacetime of interest, after Euclideanization, is~\cite{Taub:1950ez,Newman:1963yy,Page:1979aj,Page:1985bq,Page:1985hg}~\footnote{See also the discussion in ref.\cite{Stephani:2003tm}.}:
  \begin{eqnarray}
  \label{eq:metric}
ds^2 = Y( r)(d\tau+2n\cos\theta d\phi)^2
+ Y(r)^{-1}dr^2 + (r^2-n^2 )(d\theta^2+\sin^2\theta d\phi^2)\ , 
\end{eqnarray}
with
\begin{eqnarray}
Y( r) \equiv \frac{(r^2+n^2)-2Mr+l^{-2}(r^2-6n^2r^2-3n^4)}{r^2-n^2}\ , \nonumber
\end{eqnarray}
where  $\tau$ is Euclidean time with period $\beta$ and $r$ is a radial coordinate. The angles $\theta$ and $\phi$ are the standard ones on a round $S^2$.

Here, $n$ is the ``NUT" charge. If it vanishes, this spacetime is simply Euclideanized Schwarzschild (in AdS), the parameter $M$ is the mass, and the usual thermodynamic results follow by applying the standard techniques. If $n$ is non--zero the spacetime is an AdS generalization of either Taub--NUT \cite{Taub:1950ez,Newman:1963yy} or Taub--Bolt \cite{Page:1979aj}, depending upon whether the function $Y$ vanishes at $r=n$  or at $r=r_b>n$ The latter shall be our focus.  Non--vanishing parameter $n$ defines a non--trivial fibration of the $\tau$--circle over the two--sphere parameterized by $\theta$ and $\phi$. The period,~$\beta$, of  $\tau$  is fixed (in order to remove Misner strings \cite{Misner:1963fr}) to be $4n$ times the period of $\phi$, and in order to avoid conical singularities at the poles of the sphere, the period of $\phi$ is $2\pi$. So we have  $\beta=8\pi n$, which defines the inverse temperature of the solution. In the extended thermodynamics with dynamical pressure $p=-\Lambda/8\pi= 3/8\pi l^2$, the various thermodynamic quantities turn out to be \cite{Chamblin:1998pz,Hawking:1998ct,Johnson:2014yja}:
\begin{eqnarray}
H&=& \frac{r_b^2+n^2}{2r_b}+\frac{4\pi}{3}p\left(r_b^3-6n^2 r_b - 3\frac{n^4}{r_b}\right)\equiv M\ , \nonumber\\
S&=& {4\pi n}\left[\frac{r_b^2+n^2}{2r_b}+\frac{4\pi p}{3}\left(3r_b^3 \!-\!12 n^2 r_b\!-\!\frac{3n^4}{r_b}\right)\right]\ , \nonumber\\
T&=&\frac{1}{8\pi n}\ ,\quad V=\frac{4\pi}{3}(r_b^3-3n^2 r_b) \ ,\nonumber\\
r_{b}&=&\frac{l^2}{12 n }\left(1\pm\sqrt{1-48\frac{n^2}{l^2}+144\frac{n^4}{l^4}}\right)\ ,
\label{eq:EOSb}
\end{eqnarray}
where the bolt radius $r_b$ comes in two branches, the upper (the plus choice) and the lower (the minus), and $H,S, T$ and $V$ are the enthalpy, entropy, temperature, and thermodynamic volume:
\begin{eqnarray}
T=\frac{1}{\beta}\ ,\quad S=\beta\frac{\partial I_b}{\partial \beta}-I_b\ ,\quad
V=\left.\frac{\partial H}{\partial p}\right|_S\ .
\end{eqnarray}
Here, $I_b=(4\pi n/l^2)(l^2 M+3n^2r_b-r_b^3)$ is the Euclidean action of the solution, which can be computed using the methods described in {\it e.g.} refs.~\cite{Chamblin:1998pz,Hawking:1998ct} or ref.~\cite{Emparan:1999pm,Mann:1999pc}.
The expression for $r_b$ above arose by requiring regularity of the solution at $r=r_b$, which is equivalent to setting $Y^\prime=1/2n$ there (see {\it e.g.} ref.~\cite{Chamblin:1998pz}).   The upper and lower branches for the radius $r_b$ are analogous to the large and small black hole branches seen \cite{Hawking:1982dh} for the thermodynamics of Schwarzschild in AdS. Indeed the thermodynamic phase structure is not dissimilar \cite{Johnson:2014pwa}, with the upper branch being thermodynamically favoured above a transition temperature. We will stay above that transition temperature for all of this paper, and so it is the upper branch of $r_b$ that will get our attention.

It is worth remarking here that the system~(\ref{eq:EOSb}) is arranged  in quite a different way from that which arises from Schwarzschild, and indeed other black holes. For the latter, the periodicity~$\beta$ comes from removal of conical singularities arising from the vanishing of $Y(r)$ at the horizon (at $r=r_+$), and as such gets determined by  (the $r$--derivative of) $Y(r)$ there: $\beta=4\pi/Y^\prime|_{r_+}$. For Taub--Bolt, the period is fixed by the topology of the fibration of the $\tau$--circle over~$S^2$, and so~$Y$ is not involved, only $n$, which enters the fibration through the cross--term in the metric~(\ref{eq:metric}). While the temperature formula is therefore simpler, the entropy  is {\it not} 1/4 of the area of the horizon, as it is for black holes, but instead is of a more complicated form that can be attributed to additional geometrical contributions from Misner strings \cite{Emparan:1999pm,Mann:1999pc}. So there is a trade--off in where the complexity of the systems of equations come from.  On the other hand there are similarities: the mass (enthalpy) $M$ and radius $r_b$ follow from the vanishing of $Y(r)$ at $r=r_b$ in a similar way to how the analogous quantities arise for black holes. 

For comparison, here are the thermodynamic quantities for Schwarzschild \cite{Hawking:1982dh,Kastor:2009wy}:
\begin{eqnarray}
\label{eq:black}
M&=&\frac{r_+}{2}+p\frac{4\pi}{3}r_+^3\ ,\quad S=\pi r_+^2\ ,  \nonumber \\
T&=&\frac{1}{4\pi r_+}+2 pr_+\ ,\quad V=\frac{4\pi}{3}r_+^3\ .
\label{eq:EOSbh}
\end{eqnarray}
The equation of state $p(V,T)$ follows from the temperature equation, while for Taub--Bolt it is contained in the equation for $r_b$ since that is the equation relating $r_b$,~$n$, and $l$. The content of the Taub--Bolt equation of state can be seen by developing a large $T$ expansion. Take the small $n$ expansion of the upper branch $r_b$ equation  $r_b=l^2/(6n)-2n-18{n}^{3}/{{l}^{2}}+\cdots$ and writing in terms of $T$ and $p$ one gets:
\begin{equation}
r_b={\frac {T}{2p}}-{\frac {1}{4\pi T}}-{\frac {3p}{32{\pi}^{
2}{T}^{3}}}-{\frac {3{p}^{2}}{32{\pi}^{3}{T}^{5}}}+\cdots
\end{equation}
which resembles the equation for $r_+$ one would get by expanding the black hole temperature equation in~(\ref{eq:black}):
\begin{equation}
r_+={\frac {T}{2p}}-{\frac {1}{4\pi T}}-{\frac {p}{8{\pi}^{
2}{T}^{3}}}-{\frac {{p}^{2}}{8{\pi}^{3}{T}^{5}}}+\cdots
\end{equation}
Crucially, the coefficients begin to disagree at  the third term, corresponding to the dependence on~$n$.
In a manner analogous to how a high $T$ expansion can be developed  for all black hole quantities by using the~$r_+$ expansion \cite{Johnson:2015ekr,Johnson:2015fva}, one can do the same here to any desired order for Taub--Bolt using the $r_b$ expansion:
\begin{eqnarray}
V&=&{\frac {\pi{T}^{3}}{6{p}^{3}}}-{\frac {T}{4{p}^{2}}}-{\frac 
{1}{192{\pi}^{2}{T}^{3}}}-{\frac {9p}{512{\pi}^{3}{T}^{5}}}+\cdots \nonumber
\\
S&=&{\frac {\pi{T}^{2}}{4{p}^{2}}}-\frac{1}{4{p}}-{\frac {5}{64\pi
{T}^{2}}}-{\frac {p}{64{\pi}^{2}{T}^{4}}}+\cdots \nonumber
\\
M&=&{\frac {\pi{T}^{3}}{6{p}^{2}}}-{\frac {5}{32\pi T}}-{
\frac {p}{48{\pi}^{2}{T}^{3}}}-{\frac {27{p}^{2}}{512{\pi}^{3}{T}^{5
}}}+\cdots
\label{eq:EOSbT}
\end{eqnarray}
The first line (the equation of state) starts out with the same ``ideal gas" behaviour that $D=4$ black holes have: $pV^{1/3}=(\pi/6)^{1/3} T$, the same leading correction, and then the third and fourth terms show the beginning of the differences at higher order (for Schwarzschild it would be $-1/48$ and $-1/32$ for the rational coefficients, respectively), with a similar story in the entropy expansion ($-5/64$ and $-1/64$) while the mass/enthalpy expansion starts disagreeing already at second order (the black hole has instead $-1/8$, $-1/12$, and $-3/32$ respectively).

This all makes Taub--Bolt a very interesting counterpart to the Schwarzschild black hole. Rather than being some deformation of it (by a parameter such as rotation, or charge under some additional sector like Maxwell or Born--Infeld, or {\it via} the addition of high order curvature sectors) it comes with no adjustable parameter. Instead it defines  new thermodynamic equations of state that agree  with Schwarzschild in the first few orders of the high temperature expansion.

\section{Comparison of Heat Engine Efficiencies}
\label{sec:three}
As explained in section~\ref{sec:one}, it is interesting to compare the features of holographic heat engines that can be defined with these equations of state. Presumably the  engine efficiencies (compared for the same cycles in the $(p,V)$ plane) will differ, although one can anticipate that if working in a regime where the high temperature expansions hold, the differences will be small. This is easy to see since it is possible to write exact expressions for the efficiency in some circumstances~\cite{Johnson:2016pfa}. For static black holes there is a natural engine cycle made of a rectangle in the $(p,V)$ plane (where $p$ is the vertical axis) with corners labelled $1,2,3,4$ respectively (going clockwise starting from top left). Steps 1--2 and 3--4 are isobars, while the vertical steps are isochors for which there is no heat flow as $C_V=0$. Since the First Law written in terms of the enthalpy (mass) is: 
\begin{equation}
\label{eq:FirstLawB}
dH=TdS+Vdp\ ,
\end{equation}
the heat flow along the isobars is just the mass difference and so:
\begin{equation}
\label{eq:efficiencyone}
\eta_{\rm bh}=1-\frac{M_2-M_1}{M_3-M_4}\ ,
\end{equation}
where $M_i$ is the mass evaluated at the $i$th corner. Since, in the high temperature expansion, the masses of Taub--Bolt and Schwarzschild differ already by four orders, this expression suggests that the differences are going to be small, as will be conformed below.

Along similar lines, it is easy to write an expression for Taub--Bolt's efficiency on the same shape of cycle.  Here, $C_V\neq0$, and so there are heat flows along the vertical segments as well. But we can proceed  by noting  (as done in ref. \cite{Hennigar:2017apu}) that along an isochor, the First Law presented in the form~(\ref{eq:FirstLawA}) gives the heat  flows in terms of changes in the internal energy\footnote{Note that in traditional black hole thermodynamics, where $p$ is fixed, the internal energy $U$ is simply the mass. In the extended thermodynamics that we're working in, mass is actually the enthalpy, $H=U+pV$: the energy of formation of the spacetime is included~\cite{Kastor:2009wy}.} $U$. So: 
\begin{eqnarray}
\eta_{\rm tb}=1-\frac{M_2-M_1+U_1-U_4}{M_3-M_4+U_2-U_3}
=\frac{(p_1-p_4)(V_2-V_1)}{M_2-M_4-(p_1-p_4)V_1}\ , 
\label{eq:efficiencytwo}
\end{eqnarray}
where in the last step the relation $U=M-pV$ was used to write everything in terms of $M$, $p$ and~$V$.

 Using this formula exactly is hard if the engine cycle is specified in terms of pressures and volumes at each corner. The equations~(\ref{eq:EOSb}) don't readily allow for exact evaluation of $M$ (or $U$) if $p$ and $V$ are specified, because the dependence of $V$ on $r_b$, $n$, and~$p$ is rather complex. However the large $T$ expansion~(\ref{eq:EOSbT}) makes it easier to proceed, and this is what we shall use, for a numerical exploration of the efficiency comparison. At any given corner, once a $p$ and a $V$ have been specified, the temperature $T$ can be computed by solving (numerically) the equation of state (the first equation in~(\ref{eq:EOSbT})). The mass $M$ can then be computed (using the third equation in~(\ref{eq:EOSbT})), and in this way, the efficiency of the cycle evaluated using either form of equation~(\ref{eq:efficiencytwo}). This can then be compared to the efficiency of the black hole for the same cycle, computed  using  the equivalent large $T$ expansion for $M$  truncated to the same order,  and also  (as a check) the exact formula for~$M$. 
 
 The results are as follows: Wherever we reliably explored (staying well above the transition temperature for a given pressure) on the $(p,V)$ plane,  the black hole engine's efficiency was greater than that of the Taub--Bolt, but (as expected) by a very small margin. For example, for $p_1=10$, $p_4=5$ and $V_1=5$ and $V_2=10$, the black hole heat engine has $\eta_{\rm bh}=0.4986252$ while for Taub--Bolt $\eta_{\rm tb}=0.4986198$, (after rounding), and so the difference was $5.40\times 10^{-6}$. This is consistent with the fact that  $T$ averaged about 18 over the cycle, and so the differences showed up at the predicted order given that the expansions for $M$ start to differ four orders down: $(18)^{-4}\sim 9.4\times 10^{-7}$. The cycle was moved to different parts of the $(p,V)$ plane (while staying at high enough $T$) to suggest robustness of the observation, but of course it is possible that outside the high $T$ regime the difference reverses sign.

It is also possible that this test was biased toward the black hole since the rectangular cycle  could conceivably  be better adapted to the black hole system given that  there is no heat flow for its vertical segments.  This is where  the benchmarking scheme  outlined in ref.~\cite{Chakraborty:2016ssb} proves  useful.  There, it was argued that a cycle that has no special features should be used as a means of comparing the performance of different working substances ({\it i.e.} different black holes and other spacetimes) and a {\it circle} in the $(p,V)$ plane was suggested as the ``benchmarking" cycle.

\begin{figure}[h]
\centering
\includegraphics[width=2.5in]{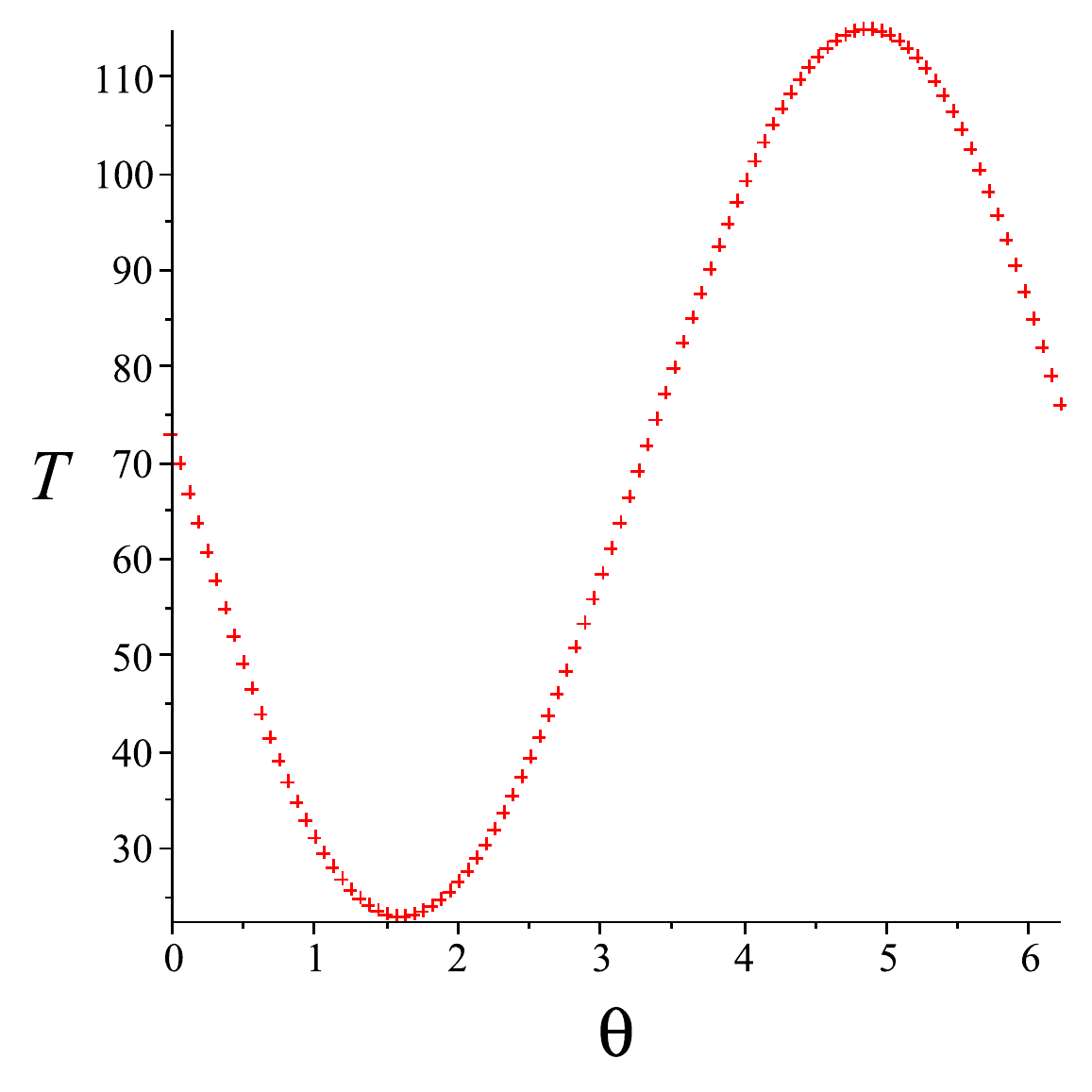} 
   \caption{\footnotesize  Temperature samples at 100 points around the circle parameterized by $\theta$ (running clockwise).}   
 \label{fig:temps}
\end{figure}

\begin{figure}[h]
\centering
\includegraphics[width=2.5in]{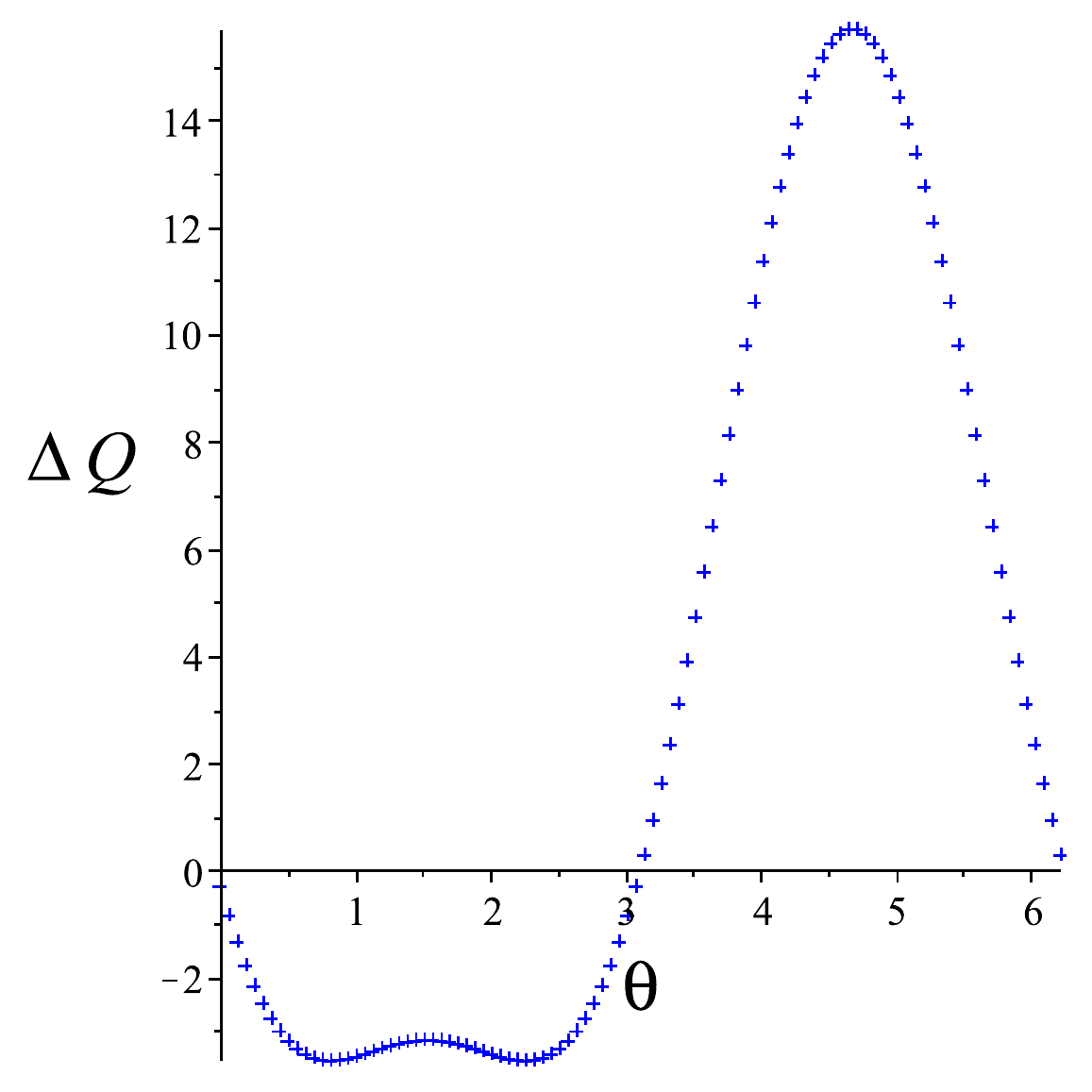} 
   \caption{\footnotesize  Heat flow samples at 100 points around the circle parameterized by $\theta$ (running clockwise).}   
 \label{fig:heats}
\end{figure}

As an example of a  benchmark circle  we chose $(p_0=15,V_0=50)$ for the circle's centre, and radius~10. The efficiency was calculated here by  a numerical evaluation of the heat flows around the circle. Figure~\ref{fig:temps} shows the temperatures computed for Taub--Bolt at 100 points around the circle parameterized by a {\it clockwise}--running $\theta$ (starting on the circle's rightmost point), while figure~\ref{fig:heats} shows samples of the heat flows at those same points. Similar results can be obtained for Schwarzschild, and the number of points were increased to 500 for added accuracy. The results for this circle were  $\eta_{\rm bh}=0.6872429297$ and $\eta_{\rm tb}=0.6872429002$, with the black hole heat engine again winning by a narrow margin ($2.95\times10^{-8}$)  as before. Benchmarking circles were computed at a variety of (sufficiently high~$T$) points on the $(p,V)$ to check robustness of this. Again, it is possible that outside the high temperature regime the difference reverses sign, but to check that would require different methods.

\section{Concluding Remarks}
It was shown in section~\ref{sec:two}, while the organizations of the thermodynamic quantities of Taub--Bolt--AdS and Schwarzschild--AdS are quite different (when written in terms of metric parameters), and while the spacetimes themselves are globally very different, closer examination reveals a great deal of (local) similarity, as shown by the form of the equations of state in a high temperature expansion. Indeed, the first two orders of the expansion for $S$ and $V$ show agreement, with coefficients beginning to deviate at further subleading orders. The enthalpies (masses) only agree at leading order, interestingly.

It was also natural to compare, as done in section~\ref{sec:three}, the two systems' behaviour as heat engines, and here a non--black hole spacetime, Taub--bolt--AdS, was used to define a heat engine for the first time. Numerical results show that the Schwarzschild--AdS equation of state  seems to define a more efficient heat engine, at least in the high temperature expansion. The efficiency is only slightly greater however, an observation  that  is traceable to the  fact that the thermodynamic quantities from which the efficiency formula is built  only differ by small amounts in this high temperature domain. An  interesting question for further work would be to see if this result persists  beyond the high temperature regime.

\medskip
 
 \section*{Acknowledgments}
CVJ  thanks the  US Department of Energy for support under grant {\rm DE--SC0011687},  the Simons Foundation for  a Simons Fellowship (2017), and Amelia for her support and patience.

\bibliographystyle{utphys}
\bibliography{johnson_taub_bolt}

\end{document}